\documentclass[12pt]{article}
\usepackage{geometry}
\usepackage{lineno,hyperref}
\usepackage[utf8]{inputenc} 
\usepackage{graphicx} 
\usepackage{amsmath} 
\usepackage{color,soul} 
\usepackage{rotating,tabularx} 
\usepackage{cite} 
\usepackage{comment}
\usepackage[T1]{fontenc}
\usepackage{authblk}
\usepackage{adjustbox}

\title{A glance into the evolution of template-free protein structure prediction methodologies}
\date{}
\author[1]{Surbhi Dhingra}
\author[2]{Ramanathan Sowdhamini}
\author[3,4,5]{Frédéric Cadet}
\author[1]{Bernard Offmann\thanks{corresponding author : bernard.offmann@univ-nantes.fr}}
\affil[1]{Université de Nantes, CNRS, UFIP, UMR6286, F-44000 Nantes, France}
\affil[2]{Computational Approaches to Protein Science (CAPS), National Centre for Biological Sciences (NCBS), Tata Institute for Fundamental Research (TIFR), Bangalore 560-065, India}
\affil[3]{University of Paris, BIGR—Biologie Intégrée du Globule Rouge, Inserm, UMR\_S1134, Paris F-75015, France}
\affil[4]{Laboratory of Excellence GR-Ex, Boulevard du Montparnasse, Paris F-75015, France}
\affil[5]{DSIMB, UMR\_S1134, BIGR, Inserm, Faculty of Sciences and Technology, University of La Reunion, Saint-Denis F-97715, France}

\begin{document}
\maketitle

\section*{Abstract}
Prediction of protein structures using computational approaches has been explored for over two decades, paving a way for more focused research and development of algorithms in comparative modelling, \textit{ab intio} modelling and structure refinement protocols. A tremendous success has been witnessed in template-based modelling protocols, whereas strategies that involve template-free modelling still lag behind, specifically for larger proteins (> 150 a.a.). Various improvements have been observed in \textit{ab initio} protein structure prediction methodologies overtime, with recent ones attributed to the usage of deep learning approaches to construct protein backbone structure from its amino acid sequence. 
This review highlights the major strategies undertaken for template-free modelling of protein structures while discussing few tools developed under each strategy. It will also briefly comment on the progress observed in the field of \textit{ab initio} modelling of proteins over the course of time as seen through the evolution of CASP platform.\\
\\
This paper is dedicated to the memory of Anna Tramontano (1957-2017) who was an Italian computational biologist and chair professor of biochemistry at the Sapienza University of Rome.\\
\\
Declarations of interest: none
\\
Keywords: protein structure prediction; ab-initio; template-free modelling; artificial intelligence\\
\\
Abbreviations: Critical Assessment of protein Structure Prediction (CASP), Template-based modelling (TBM), Template-free modelling (TBM), Fragment-based approaches (FBA), Artificical Intelligence (AI).

\section{Introduction}

Proteins are complex biomolecules that play a crucial role in building, strengthening, maintaining, protecting and repairing a living entity. Each protein folds into a specific three-dimensional structure owing to its amino acid composition. This in turn corresponds to a specific function, collectively termed as sequence-structure-function paradigm \cite{Whisstock2003}. The relationship between protein sequence and its corresponding secondary and tertiary structure is termed as second genetic code \cite{Kolata1986TryingTC}. A major gap exists in our knowledge of the science behind protein folding based on its sequence. Research focused in deciphering the second genetic code has been budding for past few decades by means of various schemes.

Advent of genomics has led to the availability of large deposit of sequence data online. This helps in easy classification of proteins and in approximating their functional annotation. A considerable amount of this classification is based on shared sequence similarity (and conserved domain search) between two or more sequences. Currently, UniProtKB/TrEMBL database is enriched with around 170 million sequence data \cite{Boutet2007}. Yet protein functionality remains unclear primarily due to the lack of structural description at the atomic levels. The equivalent structural database, RCSB \cite{Rose2017} (https://www.rcsb.org) documents around 160,000 structures to date belonging to well defined protein families. There is also an ever increasing gap between protein sequence and structure data availability due to considerable growth observed in sequencing techniques.

Scientific community has always relied on experimental approaches to deliver high resolution protein structures. Structural data deposited in data banks are only accountable when verified through experiments like X-Ray \cite{Perutz1960}, NMR \cite{Morelli2000} etc. Time and again these techniques have been proven to be most efficient in getting relevant spatial characterisation of a protein. On the other hand, they also have remained stagnant in terms of improvements due to being heavily restricted by time and manpower requirements\cite{Xu2013}. A recent introduction of Cryo-EM has fostered an acceleration of protein structure determination process \cite{Callaway2015}. The core of this technique lies in photographing frozen molecules to determine their structure. Nonetheless, the approach is relatively new and usually generates lower resolution structures than those benchmarked by other experimental techniques. 

Twentieth century has witnessed a blooming era for scientific community indulging in computational approaches for approximation of protein structures. Anfinsen in 1972 laid the foundation for protein structure prediction by correctly refolding ribonuclease molecule from its sequence \cite{Anfinsen1972}. As stated in the paper, ``the native conformation is determined by the totality of inter-atomic interactions and hence by the amino acid sequence in a given environment'' \cite{Anfinsen1973}. In other words, a protein attains its conformational nativity when its environment is at its lowest Gibbs free energy levels. Another statement put forward in their work was that a protein structure is only stable and functional in the environment it was chosen during natural selection. Despite knowing the physical environment requirement for folding a protein sequence, it remains a challenge to fold them into their functional form. Therefore, limiting the understanding of the sequence-structure-function paradigm \cite{Roy2010,Feig2017}. 

\begin{figure}[ht!]
	\begin{center}
	\includegraphics[width=\textwidth]{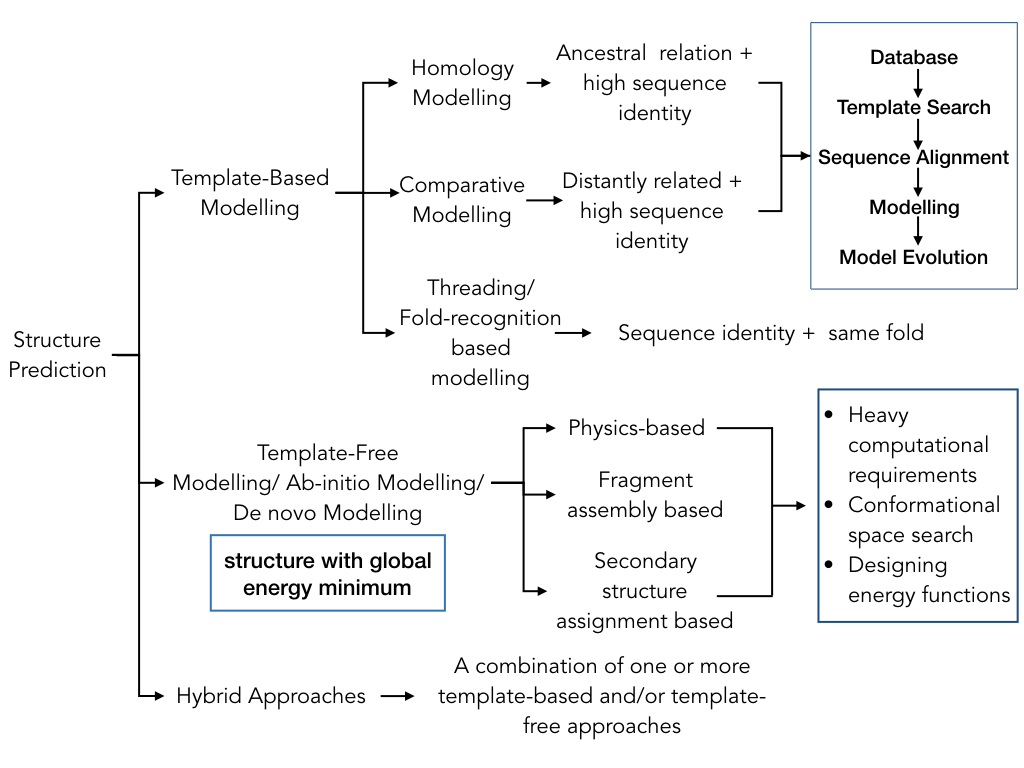}
	\caption{Computation Protein Structure Prediction Approaches.}
	\medskip
    \small
    This figure provides a broad classification of few computational protein structure prediction strategies developed and used to determine protein structure. The two major classifications of these methodologies lie under the domain of Template-Based Modelling and Template-free modelling. Each of these categories can further be split into a set of strategies based on the basic principle followed by the parent approaches for structure prediction.
	\label{fig:01}
	\end{center}
	\end{figure}

Computational approaches for protein structure prediction can broadly be categorized into two groups: Template-Based Modelling (TBM) \cite{Andras2010,Kallberg2012} and Template Free Modelling/Free-Modelling (FM)\cite{Karaka2012}. A representative flowchart of the categorization is illustrated in Figure~\ref{fig:01}. This classification has been adopted by well-known biennial competition of protein structure prediction, Critical Assessment of protein Structure Prediction (CASP) \cite{Hung}\cite{Ben-David2009,Kinch2011,Tai2014,Simoncini2017}. Results from this competition benchmark the improvement in the field of computational protein structure prediction \cite{Kinch2016,Tai2014}.
Majority of progress witnessed in this field is in construction of protein models using templates sharing high sequence similarities with unknown protein. The basis behind the approach is that similar sequence tend to fold in a similar manner. This tendency of proteins to envelop into similar folds reduces with shared sequence similarity, though there exist cases of proteins having same folds even when their shared sequence similarity is low.

TBM, as the name suggests, makes use of template to predict 3D models. Single or multiple homologous protein sharing high sequence similarity are aligned to the unknown protein sequence predicting likely models \cite{Andras2010}. Structures predicted through TBM usually have a good resolution and might fall into same functional classes. But there is little progress made when it comes to predicting new protein folds or structures. TBM is an effective approach as long as the query shares at-least 30\% sequence identity with the template \cite{Khor2015}. On the basis of shared sequence identity, it can be classified into Homology Modelling (HM) \cite{Arnold2006,Bordoli2008,Monzon2017}, Comparative Modelling (CM) \cite{Fiser2001,webb2016comparative} and Threading approaches (fold-recognition) \cite{Rost1997,Skolnick2001a,Taylor2004,Xu2008}. Each sub-class follows similar methodology into prediction of protein three-dimensional organisation from its primary one-dimensional sequence. One might argue that HM and CM are two terms for one and the same approach. It is true to a great extent except that homology modelling is defined when template shares an ancestry with the query being modelled whereas in case of CM, the query sequence has no identified evolutionary relationship with the template but only shared sequence similarity. So far, comparative modelling has been the most successful computational protein structure prediction approach available \cite{Khor2015}. The third category of TBM is fold-recognition/threading which follows the idea of picking template structures based on their fit with the protein sequence in question. It is basically a comparison of 1D protein sequence to 3D template structure.

\section{\textit{Ab initio} Protein Structure Prediction} 

A significant amount of sequence data does not share homology with well-studied protein families. This called for development of approaches which could help predicting protein structures with minimal or no known information. Such approaches fall into the second major class of computational protein structure prediction called ``Template-Free modelling/Free-modelling'' (TFM/FM). The word ``free'' used in the name indicates the initial take on such algorithms to rely on physical laws to determine protein structures. Though, most of the algorithms developed around it are guided by structural information. In this review  we will touch into the evolution of Free-modelling and the approaches that have been used to predict 3D models. Throughout this review Free-modelling, \textit{ab initio} modelling and \textit{de novo} modelling will be used interchangeably to discuss template-free modelling approaches. 

Template-free modelling comprises of algorithms/pipeline/methods for generating protein models with no known structural homologs available. Mainly these approaches focused on using physics based principles and energy terms to model proteins. The nomenclature remains debatable as in several cases, information from known structures is used in one way or another. This review is considering the following definition as best suited to describe our understanding of TFM: ``\textit{Ab initio} protein structure prediction or Free modelling (FM) can most appropriately be defined as an effort to construct 3D structure without using homologous proteins as template'' \cite{Kim2009,Kallberg2012,Zhang2015,Khor2015,Bhattacharya2016a}. FM approaches majorly depend on designing algorithms with ability to rapidly locate global energy minimum and a scoring function capable of selecting best available conformation from the several generated models\cite{Bakker2006,AdamLiwoCezaryCzaplewskiStanisawOdziej2008,Jayaram2014}.

The aim of free-modelling protocols is to predict the most stable protein spatial arrangement with lowest free energy. The major challenge faced while developing \textit{ab initio} approaches is searching conformational space which is usually huge considering the dynamic nature of proteins. Since, these approaches involve building the protein structure from scratch, focus is laid on building effective energy functions to minimise conformational search space and facilitate accurate folding \cite{Khor2015,Bhattacharya2016a}. \textit{Ab initio} algorithms can also be influenced from experimental data available in the form of abstract NMR restraints, predicted residue-residue contact maps, Cryo-EM density maps etc. \cite{Bowers2000,Topf2006,Brian2014}.

\begin{sidewaystable}
\centering
    \caption[Short Heading]{Strategies available for protein structure prediction.
    First two columns provide the name and a brief description of few algorithms developed overtime for \textit{ab inito} protein structure prediction. Third column states the length which for most tools is indicative state-of-the-art results observed throughout the field,i.e, able to achieve a GDT\_TS score of 30\% and above. For most part high accuracy models generation is still limited to small protein (< 150 residue long).}
    \label{Tab:01}
    \begin{tabular}{p{3cm}|p{12cm}|p{2cm}|p{2.5cm}}
\hline
\textbf{Algorithms/ Servers} & \textbf{Strategy/Approach} & \textbf{Maximum predicted length} & \textbf{Reference} \\
\hline
Rosetta & Fragment-assembly using MC simulations, all-atom energy function to determine structure and clustering of models & $\sim$ 250 & \cite{Bowie1994,Simons1997,Simons1999,Bonneau2001a,Raman2009,Park2015,Ovchinnikov2015,Alford2017} \\

Quark & Fragment-based assembly using REMC simulation guided by knowledge based potentials & upto 150 & \cite{Xu2012,Xu2013} \\

EdaRose & Utilising EdaFold algorithm for fragment based assembly with cluster based and energy based variations & upto 150 & \cite{Simoncini2017}   \\

Chunk-Taser & Hybrid approach using restraints derived from super secondary structure chunks as well as by threading the templates & < 200 &  \cite{Zhou2007}  \\

UNRES & Physics-based conformational space search using UNRES energy function & < 80 & \cite{Odziej2005} \\

BCL::FOLD  & Assembling secondary structural elements using MC sampling and knowledge-based energy functions & $\sim$ 200 &  \cite{Karaka2012,Brian2014,Fischer2016} \\

SS-Thread  & Prediction of contacting pairs of $\alpha$-helix and $\beta$-strand & < 200 & \cite{Maurice2014} \\

UniCon3D & Using foldons and probabilistic models to capture local backbone structural preference and side chain conformation search space & $\sim$ 200 & \cite{Bhattacharya2016a} \\

Bhageerath-H & Hybrid approach involving combination of several tools developed in the lab with the goal to reduce conformational search space & $\sim$ 350 & \cite{Jayaram2014} \\

SmotifTF & Fragment-Based approach developed using saturated library of super secondary structure fragments & $\sim$ 150 & \cite{Vallat2015} \\

Touchstone  & Secondary and tertiary restraints prediction through threading-based approaches & $\sim$ 150 & \cite{Kihara2001,Zhang2003} \\

PconsFold & Evolutionary based structure prediction pipeline using PconsC contact predictions on Rosetta folding protocols & $\sim$ 250 & \cite{Michel2014}  \\

EdaFold & Fragment-based all atom energy function to produce atomic models & upto 150 & \cite{Simoncini2012,Simoncini2013} \\

Astro-Fold & folding amino acid sequences using first-principle based approaches & $\sim$ 150 & \cite{Klepeis2003,Subramani2012} \\

DMFold & Deep learning based protein modelling by incorporating predicted structural constraints like inter-atomic distance bonds, hydrogen bonds and torsion angles & upto 200 & \cite{Greener2019}  \\

\hline
\end{tabular}
\end{sidewaystable}

\subsection{Strategies for \textit{Ab initio} Prediction of Protein Structure}

Free modelling has witnessed a major bloom in the past era owing to several strategies developed for structure prediction, few of them have been stated in the Table \ref{Tab:01}\vspace*{1pt}. Initially the scientific community resorted to use pure physics based laws, MD simulations etc. to explore the atomic dynamics of protein molecules. The prediction horizon expanded with time into utilizing restraints like C$\alpha$-C$\alpha$ distance, dihedral angles, solvent interactions, side-chain atoms, contact map information and more from available structures. The newer fundamentals involved building saturated library of structural information in the form of small fragments, secondary structural elements, motifs, foldons etc. Below we have broadly classified the \textit{ab initio} protein structure prediction approaches based on the core methodologies used to develop them.   

\subsection{Physics Based Approaches}

These formed the basis of initial algorithms built under the emerging field. The main idea behind developing these physics-based approaches is to rely on MD simulations to trace the folding path of the proteins. The philosophy backing their design is to obtain lowest energy conformation model by folding the protein sequence using quantum mechanism and coulomb potential \cite{Liwo1999,Hardin2002}. But due to high computational requirements, the field majorly relies on inter atomic interactions and force fields to solve the protein folding problem.

Free energy calculations have been explored from the very beginning of computational protein structure prediction evolution. It is believed that these approaches can go beyond documented structures and capture novel folds and patterns by exploring the inherent dynamic motion of proteins \cite{Perez2016,Raval2016}. Despite the availability of better computing, physics based approach continues to lag behind due to the amount of time required to reach the native state along-with the meddling of erroneous force-field that restrict the model to attain it \cite{Bonneau2001,Simmerling2002,Raval2012,Feig2017}.    

MELD (Modelling Employing Limited Data) \cite{Perez2016} is a recently developed physics-based protein structure prediction approach which uses Bayesian law to tap into atomic molecular dynamics of proteins for structural modelling. It has proven to be effective in determining high resolution structures of proteins up to 260 residues long \cite{Perez2016}. Similar effort was made by David Shaw's group where they utilised different sets of restraints to reduce the MD simulation runs and prevent the model from getting trapped in non-native energy state \cite{Raval2016a}. H. Nguyen et al demonstrated that the combination of an implicit solvent and a force field can result in near-native models in-case of small proteins (less than 100 amino acids) \cite{Nguyen2014}. Another group showed that simulation time can be reduced and energy landscapes can be managed using residue-specific force field (RSFF1) in explicit solvent and Replica exchange molecular dynamics (REMD) \cite{Jiang2014}. 

\subsection{Fragment Based Approaches (FBA)}

It is by far the most successful strategy used for template-free prediction of protein structure. This approach revolves around the construction of fragment libraries of varied lengths, where each fragment represents a pseudo-structure. The idea is to map information from protein fragments instead of using entire templates for constructing protein model. Segments of query sequences are replaced by the fragment coordinates recorded in the fragment library or by its predicted fold. 
Since, it is computationally exhaustive to go through all possible protein fold conformations for a structure built from scratch, fragmenting the sequence limits the number of folding patterns thus reducing the computational expense. Bowie and Eisenberg introduced Fragment-Based assembly approach to predict protein structures \cite{Bowie1994}. They used fragments of length 9 to 25 from a database of known proteins and an energy function (composed of 6 terms) that can guide building of energetically stable models \cite{Bowie1994}. This attempt set path for the evolution of computational 3D-modelling of protein structures using fragments.

Through the years several fragment-based approaches have been developed; few of which have done exceedingly well and remain the best options for \textit{ab initio} protein structure prediction to date. The basic idea behind these algorithms remains the same and typically varies with fragment type, length and scoring functions used to generate energetically minimised stable structure. Rosetta \cite{Simons1999,Bonneau2001a}, one of the most renowned fragment based approach, uses fragment libraries of length 3 and 9. It follows a Monte Carlo simulation based strategy to predict globally minimised protein models. The scoring function used in Rosetta is based on Bayesian separation of total energy into individual components. Its highest achievement has been noted during CASP11, where it correctly predicted the structure of a 256 amino acid long sequence \cite{Kinch2015}.

SmotifsTF \cite{Vallat2015} produces library of supersecondary structure fragments known as Smotifs to built probable models. The fragment library construction and utilisation is based on fragment assembly protocols. The fragment collection is governed by weak sequence similarities generating fragments on average of 25 residues in length. QUARK \cite{Xu2012} has more dynamic fragment length range of up to 20 residues which are assembled using replica-exchange Monte Carlo simulations guided by knowledge-based force-field. It has also been ranked as the best predictor in FM category for both CASP9 \cite{Kinch2011} and CASP10 \cite{Tai2014} competition and was among the two dominant tools in CASP11 \cite{Kinch2015}.

The energy functions or scoring functions used in FBA are directed by micro-state interactions existing within known protein structures. These energy terms or functions are also termed as ``Knowledge Based Potentials'' \cite{Ferrada2009}. FBA algorithms sought out to optimize these energy functions. Though and on one hand, FBA based algorithms have witnessed the most success in the biannual CASP competition by designing algorithms around the principle that certain local structures are favoured by local amino acid sequence. On the other hand, this limits their ability to search for alternate conformation of the proteins within a single run which reduces their probability of discovering a novel fold.

\subsection{Secondary Structural Elements Based Approaches}

Algorithms employing the use of SSEs for building protein models usually focus on assembling the core backbone of the protein with an exception of loop regions leading to model refinement protocols. BCL::FOLD \cite{Karaka2012} is one of such algorithms built with the objective to overcome the size and complexity limits faced by most approaches. In the later edition, restraints recovered from sparse NMR data were also incorporated in the pipeline aiding in rapid identification of protein topology \cite{Brian2014}. This was benchmarked on protein data set upto a length of 565 containing both soluble and membrane proteins. The algorithm was tested on 20 CASP11 targets, out of which it was able to produce a GDT\_TS score of 30\% on average for twelve \cite{Fischer2016}. The average GDT\_TS score was accounted for 36\%. The study was conducted by using targets belonging to different categories offered by CASP, for example T0 (regular targets), TP (predicted residue-residue contacts) and TS (NMR-NOE restraints) etc. This study also pointed out that better structures were predicted for proteins dominated by $\alpha$-helix than $\beta$-strands. The prediction accuracy also decreased with the size of the protein. 

Another algorithm based on the similar principle is SSThread \cite{Maurice2014}. It predicts contacting pairs of $\alpha$-helices and $\beta$-strands from experimental structures, secondary structure prediction and contact map predictions. The overlapping pairs are then assembled into a core structure leading to the prediction of loop regions. The contact pairing strategy employed by SSThread has been shown to be better in predicting $\beta$-strand pairs then all $\alpha$ pairs. 

\subsection{Deep Learning Based Approaches}

Quite recently neural network based deep learning approaches have seen a boom in the field of protein structure prediction. 

So far, deep learning approaches for PSP have vastly been used as one of the component in the entire pipeline rather than implicitly being implemented as the driving force. Majority of its use revolve around prediction of residue-residue contacts, which are primarily derived through co-evolutionary approaches and/or by building sequence alignment profiles \cite{Schaarschmidt2017, Wang2017,Kandathil2019}.

Recent work done by Al Quraishi \cite{Alquraishi2019} focused on building a pure deep learning based prediction approach. It was designed as a one step algorithm for prediction of protein structure relying on end-to-end deferential deep learning strategy. The emphasis was laid on not using any co-evolutionary data or information from existing templates for protein model construction. Instead, the algorithm relied on data derived solely from protein sequence in question and evolutionary profile of individual residues within the sequence. This method achieved state-of-the-art results as observed in case of \textit{ab initio modelling} protocols.

Another tool that has shown prominence in CASP13 is DeepMind's AlphaFold \cite{Senior}. It uses a two step process for protein structure determination, which also involves the use of co-evolutionary profiles to guide model building. Through this methodology high-accuracy models were constructed for 24 out of 43 test proteins achieving a TM-score of 0.7 and above in the template-free modelling domain. 

The community is still beginning to explore the benefits of  deep learning approaches into PSP. The major step back that can be encountered by these techniques would be related to lack of availability of structural data. Since, these approaches are based on training the algorithm based on the certain patterns followed by available data. Structure prediction field has always been slower than the sequence equivalent which translates into lower availability of data that can be used to train the algorithm. Hence, though deep learning based approaches can be better implemented in sequence domain of protein biology, it will take other advances in structural biology to push forward deep learning based approaches. Other problem that such approaches can be prone to is over training of the algorithms.

\section{Hybrid Approaches}

With the advancements made in computational approaches to protein structure prediction, the line between individual methodology is diminishing. Now the structure prediction community is moving forward towards the use of ``Hybrid Approaches'', which do not strictly rely on pure template based or template-free prediction criteria but on the amalgamation of both. Bhageerath \cite{Jayaram2012} is one such homology/\textit{ab initio} hybrid protocol. It is available in the form of a web-server called Bhageerath-H \cite{Jayaram2014}. The main focus of the pipeline is on reduction of conformational search space. Out of thousands of predicted models, top 5 are selected based on physio-chemical metric (pcSM) scoring function (specific to this algorithm). Efficiency of this software was put to test by using CASP10 targets with promising prediction results. After the assessment of its shortcomings, an updated version was presented in CASP12 meeting as BhageerathH+\cite{Rahul2016}.

In another study, Quark \cite{Xu2012} and fragment-guided molecular dynamic (FG-MD) were added to I-Tasser pipeline \cite{Roy2010,Yang2015} to improve on the existing protocol \cite{Zhang2015,Xu2011}. The basic idea was to introduce \textit{ab initio} generated structures from QUARK into LOMETS \cite{Wu2007} to find any hit with existing homologous template with a good TM-score. Top hits are then passed into I-Tasser pipeline for atomic refinement to obtain a structure with low rmsd. This combination produced better results for FM targets in CASP10 and CASP11 experiments than QUARK alone \cite{Zhang2014,Zhang2015}. MULTICOM\_NOVEL approach is one more example of a hybrid algorithm which was constructed by combining various complementary structure prediction pipelines including MULTICOM server, I-Tasser, RaptorX \cite{Kallberg2012}, Rosetta etc.

Chunk-Tasser can also be put into this category as it utilizes both chunks of folded secondary structural fragments along with fold-recognition to assemble protein structures \cite{Zhou2007}.

On similar grounds, an initiative was undertaken in 2014 to combine methods of the best known protein structure prediction techniques and to come up with a pipeline which could generate better structures. This initiative came to be known as WeFold, where 13 labs collaborated to merge their algorithms forming 5 major branches \cite{Khoury2014}. The outcome was promising and the authors of this study discussed on further improvements to be made in prediction protocols as a result of this 'coopetition' \cite{Khoury2014}. 

\section{Evolution of CASP and its contribution}

CASP has been a contributing factor for the work done in the field of computational protein structure prediction. It is a biennial competition being conducted for around two decades serving as a platform to judge the accuracy of prediction pipelines. It has grown overtime into a protein structure prediction platform to qualify prediction strategies coming under domains like template-based modelling, template-free modelling, refinement protocols, contact prediction etc. \cite{Jauch2007,Ben-David2009,Feig2017,Moult2018}.

To keep a track of advancement in PSP techniques, CASP prepares a list of unpredicted protein sequences in each category every two years. This provides an uniformity in assessing the advancement perceived in each area of structure prediction. The protein sequence list provided for blind testing of \textit{ab initio} modelling approaches often constitutes of protein sequences with ``soon to be released'' structures. Best models are determined on the basis of several criteria, one of them being a local-global alignment score called GDT\_TS score (Global Distance Test) \cite{Zemla2003}. It calculates the C$\alpha$ distance between residues from model and template protein at defined rmsd cut-off values. Henceforth determining both local and global similarities between two protein molecules. 

The initial achievement in protein tertiary structure prediction was observed in CASP4, but mainly for small proteins ($\leq$ 120 residues). In later years, the \textit{ab initio} prediction field remained stagnant for about a decade until the introduction of better contact prediction approaches in CASP11 competing pipelines with promising improvements in prediction accuracy \cite{Moult2016}. Similar trend was observed in CASP12 with the inclusion of alignment-based contact prediction methods \cite{Abriata2018}.

Recently conducted CASP13 demonstrated further improvement on average GDT\_TS score due to the employment of deep learning approaches in structure prediction \cite{Alquraishi2019}. This served as an encouragement to dig further into deep learning based approaches to solve the protein folding problem.

\section{Conclusion}

Template-based prediction in general are quicker than experimental methods, at least in providing initial spatial arrangement of the protein. One of the major drawback of these approaches is the redundancy of information, i.e., no new fold or family can be discovered as it relies on building models from existing structures. In addition, these methods fail to establish the structural integrity of a protein sequence with decreasing sequence or structure identity.

This review peeks into few methods and possibilities of free-modelling techniques developed and available for the prediction of protein structure. \textit{Ab initio} protein structure prediction still bare influence from PDB structures for optimizing the parameters of protein folding. This information helps them reduce the conformational space sampling requirements by maximizing the efficiency of energy functions. Most of the algorithms are still directed by a combination of knowledge-based potentials and physics-based approaches \cite{Zhang2008}.

To date free-modelling has been been well adapted for protein sequences length of 150 residues or below \cite{Hung} \cite{Zhang2008,Lee2009}. Few instances have seen algorithms overdoing themselves and going beyond the length restrictions to predict structure for longer proteins. CASP11 witnessed major success in \textit{ab initio} protein structure prediction for a structure of length 256 a.a. \cite{Moult2016}. 

The major challenges faced by this field starts with finding an efficient way to explore the conformational search space as a protein sequence can fold into indefinite forms. Thereby, reducing the plausible folding possibilities to the best probable fold is a hard task to achieve.

The length limitation could indicate towards the design strategy of the algorithms, many of which rely on defining domain boundaries prior to structure prediction. But, this would not be a very strong case of argument as most reliable predictions still lie under the length of 150 residue, though single domain boundary could expand upto a length 200 to 250 residues. The only exception to this case seen over the past years of CASP competition where a protein of length 256 a.a. was accurately predicted \cite{Moult2016}.

One could also argue that as a community, we are still at the domain-level of structure prediction given the length limitations. Algorithms have been built that specifically target solving the structure of single-domain proteins \cite{Kim2004,Michel2014} or choices of restraints that are limited to single-domain proteins only \cite{Zhang2003}. Many of the designed algorithms tend to be validated on a data set of single-domain proteins as well \cite{Schaarschmidt2017}. 

Even so, in the case of small protein structure predictions, not all of them deliver model close to its nativity. It is just that chances of having a good structure is more if the length of the protein is around 150 or less.

There is also the case of stagnancy in the field for the majority of years and been pointed out a lot. One of the example is the discussion provided by the end of CASP10 \cite{Kryshtafovych2014}. The study pointed out that the results from CASP10 lay closer to what were observed during CASP5. It also reflected on the fact that it might be due to the gradual increase in the complexity of CASP targets along with the inclusion of multi-domain targets provided for modelling.

Another drawback faced by few of these algorithms can be the run time which can vary a lot depending on the size of the protein and the internal functioning of the algorithm. With the inclusion of artificial intelligence, the time scale for modelling has been reduced to milliseconds but generally, a single prediction can take from somewhere from few minutes to hours to days for an algorithm to complete. 

Most of the algorithms still rely on manual intervention to complete the runs and so the human error should also be considered.

The point of PSP is not just high accuracy structure determination but also to ascertain the basis behind this biological process (protein folding). Thereby, finally answering questions like ``why good protein become faulty and cause disease''.

\textit{De novo} protein structure prediction still requires a lot of improvement, but at the same time it promises a better prospect of structure prediction in future. It brings with it a hope of predicting newer folds at a faster pace when compared to experimental approaches which can remain stuck for years altogether due to numerous reasons. In general computational structure prediction techniques though have a room for improvement are still quick when compared to traditional approaches \cite{Hung}. If considering Template-Based modelling approaches, few limitations still persist whereas \textit{ab initio} approaches can move a step ahead and might help understanding the basic principles of protein folding \cite{Lee2009}.

\section*{Acknowledgements}
The authors are most thankful to Yves-Henri Sanejouand for critical reading of the manuscript. SD is thankful to Conseil Régional de La Réunion and Fonds Social Européen for providing a PhD scholarship under tier 234275, convention DIRED/20161451. BO is thankful to Conseil Régional Pays de la Loire for support in the framework of GRIOTE grant.

\section*{Conflict of interest}
Authors declare no competing interests.

\section*{Authors contributions}
Conception of the work was done by SD, RS, FC and BO. SD, RS and BO performed literature data collection and analysis. All four authors SD, RS, FC and BO wrote the manuscript.\\
All authors approved the final version of the article. 

\bibliographystyle{unsrt}
\bibliography{review_dhingra_revised_v2}

\end{document}